\documentclass[11pt]{article}
\usepackage{graphicx}
\usepackage{amssymb,amsmath}
\usepackage{epstopdf}
\title{Mathematical Structure of Tetrad\\ Equations for Vacuum Relativity}
\author{Frank B. Estabrook\\
Jet Propulsion Laboratory, California Institute of Technology\\
4800 Oak Grove Drive, Pasadena, CA 91109}
\begin{document}
\maketitle
\abstract The tetrad partial differential equations formulated by Buchman and Bardeen for vacuum gravity are shown to be well posed by calculation of the Cartan characters of an associated exterior differential system.  Gauge specializations are discussed.  A Cartan 4-form is found for this field theory, together with its intrinsic version the Lagrangian density. 
\section{Introduction} General relativity was discovered and first understood as elegant Riemannian differential geometry.  For vacuum, this nonlinear theory of gravitation is also a well posed field theory. Expressing it most concisely, Buchman and Bardeen \cite{BB} have formulated and tested the tetrad equations of vacuum general relativity as a first order set for numerical integration.  Their dependent variables are first the connection variables of Estabrook and Wahlquist \cite{ERW}\cite{EW64}, a partition of the 24 Ricci Rotation Coefficients $\Gamma_{ijk}$ ($=-\Gamma_{ikj}), \mbox{where }i,j,k=1...4$ into 3-dimensional arrays that respect metric signature and local physics and that satisfy directional derivative evolution and constraint equations that are symmetric and hyperbolic (the so-called ``dyadics" $K_{ab},N_{ab},\omega_a,a_a,\mbox{where }a,b=1...3$). (An alternate partition of the $\Gamma_{ijk}$ is the set of 24 Newman-Penrose spinor components.)

Buchman and Bardeen go on to introduce 16 contravariant coordinate components of the orthonormal tetrad field, which we will write $_i\lambda^j$, so that partial derivative equations (pde's) can be explicitly written.  They in fact place both indices to the right, as $\lambda_i^j$, using Greek indices for the range  0,1,2,3.  The former notation was used in [2] and [18] and intervening papers to clearly distinguish the left index as a label of the tetrad legs (in the terminology of Synge, who put such indices in parentheses, it is a ``Lorentz" index), while the right is a customary coordinate (or tensor) index that is contravariant or covariant, raised or lowered with a metric tensor that is a function of position. A Lorentz index is also conveniently raised or lowered, using a (usually diagonal) matrix of constants coding the signature of the space.  (When there are only Lorentz indices and no chance of ambiguity, as in the  $\Gamma_{ijk}$ themselves, the indices revert to the right side.)  In the following we initially take the $_i\lambda^j$ (actually, its transposed inverse) as simply a matrix of 16 dependent field variables, with i and j labeling rows and columns. Both the tetrad orthonormality properties and coordinate covariance properties, the metric interpretations of the indices, emerge as a result of the partial differential equations. Buchman and Bardeen's partition of these additional fields is a set of dyadic arrays denoted by them $B_a^i,A_b,\beta^i,\alpha$, with spatial indices that range over 1,2,3. The last two of these are shift and lapse, respectively. Their final result is a set of 30 symmetrizeable hyperbolic evolution equations for a total of 40 dependent dyadic fields as functions of 4 independent variables $x^a, t$, together with 20 further ``constraint" equations not involving $\partial/\partial t$. For short we refer to this first order partial differential system as the WEBB equations.

Tetrad and coordinate gauge choices of 10 more evolution equations, together with consistent additional constraint equations, remain to be made.  Special orthonormal framing due to Nester \cite{nes}, adds six evolution equations for the $\Gamma_{ijk}$ and six constraints. The ``Lorentz" gauge of van Putten and Eardley \cite{vanP} adds just 6 evolution equations. Harmonic or generalized harmonic coordinate specialization \cite{pret} can add 4 final evolution equations, for a total of 40, but other choices are of course possible.

Here we use Cartan's theory of Exterior Differential Systems (EDS) \cite{ivey} to discuss the underlying mathematical structure of the WEBB equations.  After setting up an equivalent ideal of exterior differential forms in 44 dimensions, we report a calculation of its Cartan characters, a table of integers \cite{ivey, ew89}, confirming its Cauchy properties as a well posed field theory. Knowing the characters we can count the number of evolution equations, and the number of first class constraints, second class constraints, and so on.  Cartan characters are also given for different gauge choices. We then derive the underlying Cartan 4-form $\Lambda$ and its exterior derivative the multisymplectic (or Poincar\'{e}-Cartan) 5-form, and show how the WEBB system results from arbitrary variation.  They are in a deep sense Euler-Lagrange pde's, but instead of including a set of contact 1-forms containing all the partials of the potential fields (as do, e.g., the analogous EDS for source-free Maxwell \cite{betounes} or Yang-Mills \cite{hermann}) they are built on the torsion 2-forms of Cartan's movable frame bundle, involving at each point a subset of those partials. In the intrinsic version of $\Lambda$ as a Lagrangian density, the 16 $_i\lambda^j$ are potential fields, and the $\Gamma_{ijk}$ are intensity fields, combinations of partial derivatives of the potentials, and the dynamics, as expressed by field equations for Ricci-flatness, is found by variation.

\section{The Exterior Differential System}We label 44 coordinates in anticipation of the use to which we will put them as an explicit description of 4-dimensional movable frame geometry, i. e. $x^i,  ^i\lambda_j, \Gamma_{ijk}$.  The basis 1-forms in this space are thus $dx^i, d^i\lambda_j, d\Gamma_{ijk}$. In terms of them, we then define sets of 1-forms $\theta^i$ and $\omega_{ij}$ (which will shortly be seen as, respectively, tetrad and connection fields) according to
\begin{eqnarray}
    \theta^j \mbox{ }\equiv \mbox{ } ^j\lambda_k\mbox{} dx^k \nonumber\\
 \omega_{ij}\mbox{}\equiv  \mbox{}\Gamma_{kji}\mbox{ } ^k\lambda_l dx^l
\label{defs} 
\end{eqnarray}For generality in setting up the EDS we will use a matrix of constants $\eta_{ij}$ to allow for various metric signatures.  For space-time, and to ensure a hyperbolic system, $\eta_{ij}=\eta^{ij}=diag (1,1,1,-1)$.  In the following all indices \emph{except for that on $dx^i$ and the right-hand index on $^i\lambda_j$} will be raised or lowered with the $\eta$ matrix so that the $\eta_{ij}$ need not be explicitly written. Note also that we here follow the conventions of \cite{ERW} and its predecessors in the tetrad components of $\omega_{ij}$ while Buchman and Bardeen  \cite{BB} label them in the opposite order:  $\Gamma_{kji}(EW)=\Gamma_{ijk}(BB)$.

For $\theta^i$ and $\omega_{ij}$ to describe a movable frame field and a metric geometry on a submanifold of our coordinate space, $\omega_{ij}$ must be antisymmetric (it is), and the torsion 2-forms $T^i$ must there vanish:
\begin{equation}d\theta^i+\omega^i_{.j}\wedge\theta^j\equiv T^i\rightarrow0
\label{torsion}
\end{equation} 
The induced curvature or Riemann 2-forms $R_{ij}$ of the submanifold will be
\begin{equation}d\omega_{ij}+\omega_{ik}\wedge\omega^k_{.j}\equiv R_{ij}
\label{curvature}
\end{equation}
The four Ricci (better, Einstein) 3-forms (which we will want to have vanish) are then $R_{ij}\wedge\theta_k\epsilon^{ijkl}$ and the Ricci ``scalar" is the density 4-form $R_{ij}\wedge\theta_k\wedge\theta_l \epsilon^{ijkl}$.  From these definitions it is easy to calculate
\begin{eqnarray}dT^i&=&d\omega^i_{.j}\wedge\theta^j-\omega^i_{.j}\wedge d\theta^j \nonumber\\
                    &=&(R^i_{.j}-\omega^i_{.k}\wedge\omega^k_{.j})\wedge\theta^j-\omega^i_{.j}\wedge(T^j-\omega^j_{.k}\wedge\theta^k)\nonumber\\
                    &=&R^i_{.j}\wedge\theta^j \mbox{ mod } T^i
\label{dT}
\end{eqnarray}and
\begin{eqnarray}dR_{ij}&=&R_{ik}\wedge \omega^k_{.j}-\omega_i^{.k}\wedge R_{kj}\nonumber\\
d(R_{ij}\wedge\theta_k\epsilon^{ijkl})&=&0 \mbox{ mod } R_{ij}\wedge\theta_k\epsilon^{ijkl} 
\label{dR}
\end{eqnarray}

We now can set an ideal of forms (on the 44 dimensional space) whose solutions (submanifolds on which the ideal is pulled back to vanish \cite{ivey, ew89}) will be vacuum metric 4-geometries framed by four othonormal 1-forms $\theta^i$. This EDS is generated by the torsions, their closure 3-forms and the Ricci 3-forms,  and we will analyze and interpret it with Cartan theory in the next section:
\begin{equation}\{T_i, R_{ij}\wedge\theta^j, R_{ij}\wedge\theta_k \epsilon^{ijkl}\}
\label{theEDS}
\end{equation}

These generating forms, it should be emphasized, are all here being given explicitly in terms of the 44 coordinates and coordinate bases $dx^i, d^i\lambda_j$ and $d\Gamma_{ijk}$.  This ideal leads to the 50 WEBB partial differential equations when one adopts the $dx^i$ as independent variables in a solution, inserting $d\Gamma_{ijk}\rightarrow\partial\Gamma_{ijk}/\partial x^s~dx^s$ and $d^i\lambda_j\rightarrow\partial^i\lambda_j/\partial x^s~dx^s$, after which equating to zero the coefficients of the independent 2-forms $dx^i\wedge dx^j$ and 3-forms $dx^i\wedge dx^j \wedge dx^k$. In a solution the $^i\lambda_j$ become covariant components of the tetrad frame, and the $\Gamma_{ijk}$ its Ricci Rotation Coefficients.  Buchman and Bardeen more neatly write this system of partial differential equations using the equivalent transposed inverse matrix of variables $_i\lambda^j$, contravariant tetrad components related to our covariant components by   
\begin{eqnarray}\delta_i^j &=& _i\lambda^k\mbox{ } ^j\lambda_k\nonumber\\
0 &=&  _i\lambda^k\mbox{ }d^j\lambda_k +\mbox{ }^j\lambda_k\mbox{ }d_i\lambda^k
\label{inversion}
\end{eqnarray}Finally, the metric induced on the solutions is
\begin{equation}g_{ij}=\eta_{kl} \mbox{ } ^k\lambda_i \mbox{ } ^l\lambda_j
\label{metric}
\end{equation}
\section{Cartan Characters, Gauge Freedom and Constraints}  Cartan characters are integers $s_i$, $i = 0,1,2,...$ calculated from the ranks of local linear equations for the components of successive vectors $V_1, V_2, V_3$, etc., that together span nested solution manifolds of an EDS.  The $V_i$ are independent up to a limit set by the rapidly growing number of equations to be satisfied\cite{ivey}\cite{ew89}.  The characters describe the freedom of construction and the maximal dimensionality of solutions obtained by successive integrations along a flag of V's (thus giving the number of independent variables that can be chosen to write equivalent pde's);  conversely the characters describe the constraints that arise on successive nested slicings of lower dimension, starting from a generic solution \cite{ivey}. Simple tests using the characters show whether the system is well posed (or causal), distinguish allowable choices of dependent and independent (involutory) variables for writing pde's, and, among the latter isolate the the essential ones from those reflecting gauge degrees of freedom. We have calculated the Cartan characters both for the WEBB EDS Equation (6) and for versions with various gauge specializations to be given next, using Monte Carlo methods developed and programmed in Mathematica by H. D. Wahlquist\cite{ew89}.

Our notation for the Cartan characters we report places them in a conventional array or table $N\{s_0,s_1,s_2,...s_{q-1},0,0,0,...\}g$ where N is the total number of variables and the $s_i$ are the successively computed characters. We consider only "well-posed" EDS where for sufficiently high i (say q-1) the construction has no more freedom but the sum of the $s_i$ has not exceeded N-q.   So we set $g=N-\sum_{i=0}^{q-1} s_i$. It is to be understood that there are g-q zeros in the array after $s_{q-1}$.  g is called the {\it genus} by Cartan: it is the maximal dimension of an integral manifold constructed sequentially as we have described, in practice a generic solution of the associated pde system.  The zeros indicated that in fact the number of independent variables which can be taken is only q{---}the solution manifold is itself fibered over a q dimensional base, the g-q-dimensional fibers express gauge degrees of freedom. Alternatively, we can omit the zeros and report, at the last place of the table, q + fiber dimension. Additional gauge choices{---}more forms in the EDS{---}will result in a lower fiber dimension. In the limit when there is no remaining gauge freedom the solution dimension is q=g.  For all systems considered here q=4.

The principal result of this paper is the character array we have calculated for the WEBB EDS Equation (6).  We find: 44\{0,4,12,14\}4+10. The EDS passes Cartan's test for being well posed \cite{ivey}.  The number of evolution equations implied is $\sum_{i=0}^{q-1}s_i =30$, which agrees with the 10 gauge degrees of freedom.  The total number of pde's will be $\sum_{i=0}^{q-1} (q-i) s_i =50$.  The number of (first class, second class, etc.) constraints is $\sum_{i=0}^{q-2} (q-i-1) ) s_i=20$. The number of second class constraints is $\sum_{i=0}^{q-3} (q-i-2) s_i=4$.

The 10 fiber dimensions can be understood from the invariance properties of the EDS we have set, Eq. (6).  An EDS is an ideal of forms, and equivalent sets of generators are related by arbitrary invertible algebraic superposition.   Six arbitrary tetrad rotations leaving the $\eta_{ij}$ invariant are allowed due to the homogeneous structure of the Lorentz indices in Eq. (6). Four more continuous invariances (or \emph{isovectors}) are generated by vector fields $\partial/\partial x^i$, whose Lie derivation acting on the generators vanishes (the $x^i$ do not appear in the EDS except as exterior derivatives).  Evidently this is the covariance property distinguishing the $x^i$ as coordinates in the base.
\section{Gauge Specializations}The Nester tetrad gauge \cite{nes}\cite{ERW}\cite{BB} adds two closed 2-forms to the EDS:
\begin{equation}d(\Gamma^k_{ki}\theta^i) \mbox{ ,     } d(\Gamma_{ijk}\theta_l\epsilon^{ijkl})
\label{withnester}
\end{equation}
The character table computed for this EDS is 44\{0,6,14,16\}4+4.  36 evolution equations and 26 constraint equations (6 are second class).

The Lorentz gauge \cite{vanP}\cite{BB} adds 6 4-forms to the basic WEBB EDS:
\begin{equation}d(\Gamma_{ist}\theta_j\wedge\theta_k\wedge\theta_l \epsilon^{ijkl})
\label{withlorentz}
\end{equation}
The character array is 44\{0,4,12,20\}4+4.  36 evolution equations, 20 constraints (4 second class).  Since $q=4$ and the Lorentz specializations are 4-forms, their closure 5-forms will not change the character table, so in fact 6 4-forms like $f dx^1\wedge dx^2 \wedge dx^3 \wedge dx^4$, with f an arbitrary function, could have been added to them.  Buchman and Bardeen \cite{BB} recognize this in proposing various ad hoc right hand sides to the evolution equations for the components of $\Gamma_{ijk}$ that enter Eq. (9) and (10), the dyadics $\omega_a$ and $a_a$.  The left hand sides of these evolution equations are the same in both Nester and Lorentz specializations, but the right hand sides in Nester gauge come from 2-forms, are consistent with the Nester constraint equations, and cannot be arbitrarily changed.

Harmonic coordinates \cite{ERW} specialize the $^i\lambda_j$ fields with 4 4-forms:
\begin{equation}d(_i\lambda^s\theta_j\wedge\theta_k\wedge\theta_l\epsilon^{ijkl}) 
\label{withharm}
\end{equation}
Note that for conciseness we here have used the contravariant set of tetrad variables $_i\lambda^s$.  The WEBB character table including these is 44\{0,4,12,18\}4+6.  34 evolution equations, 20 constraints (4 second class).  Combined with Nester tetrad constraints one finds 44\{0,6,14,20\}4 with no gauge freedom left. 40 evolution equations and 26 constraints (6 second class).  Combined harmonic and Lorentz yields 44\{0,4,12,24\}4.  40 evolution, 20 constraints (4 second class).  Since the harmonic conditions are 4-forms the same remark as above applies, that the resulting equations can be generalized with arbitrary functions on the right hand sides of evolution equations for the shift $\beta^a$ and lapse $\alpha$. Generalized harmonic coordinates have been advocated by Pretorius \cite{pret}.
\section{Field Theory}Field theories are usually posited as Lagrange densities, and this can readily be done for the WEBB equations.  The EDS approach is to formulate a higher rank Cartan  form, here a 4-form, as a  generalization of the Cartan 1-form $\Lambda=L(q^i,\dot{q^i},t) dt$ of classical mechanical systems.  Variation of the Lagrangian $L$ yields canonical equations of motion.  Variation of the Cartan 4-form $\Lambda$ is to yield an EDS for the solution geometry and its field equations.  In the next Section we show how, given $\Lambda$ and its EDS, one can finally write a field theoretic, or intrinsic, Lagrangian density in terms of partial derivatives.

Variation of $\Lambda$ is achieved by Lie derivation along an {\it arbitrary} vector field, and requires that the exterior derivative $d\Lambda$ (a 5-form) be a sum of terms {\it each} of which is at least quadratic in the forms generating the EDS.  Then the arbitrary variation will indeed vanish on solutions. Such an exact 5-form is denoted the multisymplectic or Poincar\'{e}-Cartan form of the field theory, and is a generalization of the symplectic 2-form of classical mechanics.  For deep understanding of this approach, and bibliography beginning with Caratheodory, Le Page, Dedecker, Rund and many others, cf. \cite{GIM} \cite{BRY}.

A 5-form can be constructed that is quadratic in the torsion 2-forms and Ricci 3-forms of the basic WEBB EDS:
\begin{equation}d\Lambda\equiv T_l\wedge R_{ij}\wedge\theta_k\epsilon^{ijkl} 
\label{PC}
\end{equation}
Each term is the product of a torsion 2-form $T_l$ and a Ricci 3-form $R_{ij}  \wedge \theta_k \epsilon^{ijkl}$.  We have denoted it $d\Lambda$ anticipating that it is exact, and so a multisymplectic form for the EDS.  Indeed a quite long calculation, using Equations (4) and (5) and expanding the summations, verifies this.  $d\Lambda$ can then be integrated by parts to yield the Cartan form:
\begin{equation}\Lambda=R_{ij}\wedge\theta_k\wedge\theta_l\epsilon^{ijkl}+d\mbox{(arbitrary 3-form)}
\label{Lambda}
\end{equation}
It is satisfying that we can recognize $\Lambda$ as the Ricci scalar, the Hilbert action density for vacuum relativity.

The comment should be made that this factoring of the multisymplectic form $d\Lambda$ into a sum of products of 2-forms and 3-forms is different from that encountered in most of the literature on Cartan forms for field theories \cite{betounes,hermann,GIM,BRY}  The more usual structure is as a sum of products of 1-forms and 4-forms, with the 1-forms in the EDS being contact forms relating scalar potentials and fields and the 4-forms coding the dynamics.  Another approach to equations for vacuum gravitation that used an EDS for embedding of 4-geometry in flat 10-dimensional metric geometry \cite{ERW99} led to a Cartan form (a ``dimensionally continued" Gauss-Bonnet form \cite{MH}) whose exterior derivative is a sum of terms factorable in either of the two ways. The ideal of 1-forms, closure 2-forms and 4-forms that results from the first factoring (\emph{isometric} embedding) was shown to be well posed; the other factoring, into 2-forms and 3-forms similar to those of the present paper, was subsequently considered and is well posed with the same character table as that of the basic WEBB EDS  \cite{EST2002}.  The embedding EDS' are set on a 55 dimensional group space (the contraction of the  orthogonal group O(11) to ISO(10), rotations and translations in 10 dimensions). Explicit coordinates on the group space are not easy to introduce; also compared to WEBB there is a larger number of degrees of gauge freedom, viz. 21 (6 for O(4) tetrad rotation and 15 for O(6) dual tetrad rotation.)  The group theoretic embedding EDS may prove appropriate for global topological investigations.  Alternative well posed field theories are also suggested by the embedding formalism \cite{EST2002}.
\section{The Intrinsic Lagrangian Density} To express the variational principle in the more traditional form of a Lagrangian density dependent on potential fields and their partials, we will begin with a Cartan 4-form $\Lambda$ equivalent to the above Ricci 4-form for the WEBB EDS, viz.
\begin{equation}\Lambda=\omega_{is}\wedge\omega^s_{.j}\wedge\theta_k\wedge\theta_l \epsilon^{ijkl}-2\omega_{ij}\wedge\omega_{ks}\wedge\theta^s\wedge\theta_l \epsilon^{ijkl}+2\omega_{ij}\wedge T_k\wedge\theta_l \epsilon^{ijkl}
\label{secondLambda}
\end{equation}
This form differs from Eq. (13) only by a ``surface" term, viz., the exact 4-form $d(\omega_{ij} \wedge \theta_k \wedge \theta_l \epsilon^{ijkl})$.  Its first two terms have been given previously as a "pure grade" Cartan form in the Clifford algebra of the second frame bundle over 4-space (where there is no torsion)\cite{E91}. Equivalent formulations in other notations can be cited, e. g. \cite{Nes, Kam}. When we restrict $\Lambda$ to a 4-space, $\Lambda \rightarrow L  dx^ 1\wedge dx^2 \wedge dx^3 \wedge dx^4$, adopting the $x^i$ as independent variables, the resulting Lagrangian density is a function of $^i\lambda_j$ and $\Gamma_{ijk}$ :
\begin{equation}2L = |Det ^i\lambda_j|(\Gamma_{i..}^{.ik}\Gamma_{.jk}^{j..}-\Gamma_{ijk}\Gamma^{jik})
\label{ell}
\end{equation}
The first factor is alternatively $|Det _i\lambda^j|^{-1}$.

To set up Euler-Lagrange equations we need to vary $L$ consistently with the relations between $^i\lambda_j$ and $\Gamma_{ijk}$ in the 4-space. Setting $T^i=0$, i.e., using the EDS to restrict the action {\it in advance} to a solution 4-space, yields a combination of $\Gamma_{ijk}$'s denoted the ``Object of Anholonomity" $\Omega_{ijk}$ = $-\Omega_{ikj}$ in terms of partial derivatives of the $^i\lambda_j$:
\begin{eqnarray} 2 \Omega_{ijk}=\Gamma_{kji}-\Gamma_{jki}\\     \partial^i\lambda_p/\partial x^s-\partial^i\lambda_s/\partial x^p+2 \Omega^i_{.jk}\mbox{} ^k\lambda_s \mbox{}^j\lambda_p=0
\label{dT}
\end{eqnarray}
The set of 24 $\Omega_{ijk}$ are of course equivalent to the Ricci Rotation Coefficients $\Gamma_{ijk}$ as they can then be inversely solved for:
\begin{equation}\Gamma_{tsr} = \Omega_{str} - \Omega_{rts} + \Omega_{tsr}
\label{anholonomity}
\end{equation}
Thus in $L$ the 16 $^i\lambda_j$ will play the role of potential fields and the $\Omega_{ijk}$ or $\Gamma_{ijk}$ are intensity fields, from Eq.(17) a linear subset (``curls") of partials of the $^i\lambda_j$.  The variations of $^i\lambda_j$ and $\Gamma_{ijk}$ in $L$ must respect Eq.(17), which is 24 of the pde's in the WEBB set (12 evolution equations and 12 constraint equations for the $^i\lambda_p$). 16 more WEBB equations (12 evolution equations and 4 constraints) come from the integrability conditions  $dT_i=0$, the so-called Differential Identities satisfied by any tetrad field whether metric or not \cite{EW64}\cite{schouten}:
\begin{equation}\Omega_{r[st,p]} = 2 \Omega^q_{.[ps} \Omega_{|r|t]q}
\label{identities}
\end{equation}

We vary $L$ in the customary way using Eq.(17) to express the variations $\delta\Omega$ in terms of variations of the $^i\lambda_j$ and of the $\partial ^i\lambda_j/\partial x^k$, permute $\partial /\partial x^k$ and $\delta$ and integrate by parts, discard surface terms, finding 16 dynamic equations from the vanishing of the coefficients of $\delta^i\lambda_j$.  They are first order in the $\Omega_{ijk}$ or second order in the $^i\lambda_j$:
\begin{equation}^i\lambda_p \partial L/\partial ^j \lambda_p - 2 \Omega_{pmj} \partial L/\partial   \Omega_{pmi}+ D_m \partial L/\partial \Omega^j_{mi}=0
\label{ELeq}
\end{equation}
where
\begin{equation}D_m = _m\lambda^s \partial /\partial x^s
\label{DD}
\end{equation}
are directional derivatives in the solution.  Inserting the Lagrangian Eq.(15) six of these equations turn out to be already included in the set of 16 in Eq. (19).  The new ones are the dynamics of this field theory, viz. the final 10 WEBB equations (6 evolution, and 4 ``energy and momentum" constraints), which state in fact the vanishing of the Ricci tensor for vacuum.
\section{Acknowledgements} I thank Luisa Buchman for many useful discussions on the structure, motivation and practical use of the WEBB equations.  Lee Lindblom's insistance on seeing an explicit intrinsic Lagrangian clairified our understanding of how the Cartan form works.  This research was performed at the Jet Propulsion Laboratory, California Institute of Technology, under contract with the National Aeronautics and Space Administration, funded through the internal Research and Technology Development program.

\end{document}